\documentclass[twocolumn,showpacs,preprintnumbers,amsmath,amssymb,pra]{revtex4}
\usepackage{epsfig}

\begin{document}

\title{Virial expansion coefficients in the harmonic approximation}
\author{J.R. Armstrong, N.T. Zinner, D.V. Fedorov and A.S. Jensen }
\affiliation{ Department of Physics and Astronomy, Aarhus University, 
DK-8000 Aarhus C, Denmark}

\date{\today}
\begin{abstract}
The virial expansion method is applied within a harmonic approximation 
to an interacting $N$-body system of identical fermions. 
We compute the canonical partition functions for two and three
particles to get the two lowest orders in the expansion.  The energy
spectrum is carefully interpolated to reproduce ground state
properties at low temperature and the non-interacting large temperature limit of
constant virial coefficients.  This resembles the smearing of shell
effects in finite systems with increasing temperature. 
Numerical results are discussed 
for the second and third virial coefficients 
as function of dimension, temperature, interaction, and the 
transition temperature
between low and high energy limits.
\end{abstract}

\maketitle

\section{Introduction}

The virial expansion is a classical concept 
\cite{thie85,onne01,urse27,hill56} which has been
extended to be applicable for quantum mechanical systems \cite{kahn38}.
The expansion is in terms of few-body correlations and therefore most
efficient when the influence of $N$-body effects decrease with $N$.
In practice, this decrease has to be very fast, because higher order
correlations are extremely difficult to obtain by accurate
calculations. This fact is not obvious since only the spectrum of $N$
interacting particles is needed, not wave functions or structure nor
any other properties. However, obtaining these
spectra imply solving the $N$-body problem which is already
demanding beyond two particles for general interactions.

In the classical textbook by Huang on statistical mechanics
\cite{huang87}, the virial expansion is discussed for the 
quantum mechanical case and it is elegantly demonstrated 
how the second virial coefficient can be obtained from 
knowledge of the two-body scattering phase shift and bound state
spectrum (when present). This was subsequently generalized in the 
seminal paper of Dashen, Ma, and Bernstein \cite{dashen1969} where
a formulation of statistical mechanics in terms of the scattering 
$S$-matrix is given. The formulation gives a prescription for 
calculating virial coefficients at any order, and shortly 
thereafter the behavior of the third order coefficient 
at low temperature was obtained via the $S$-matrix method \cite{adhikari1971}. 
The $S$-matrix approach to virial coefficients is still an 
actively research topic \cite{how2010} with recent applications
in the field of cold atomic gases \cite{leclair2012}. However, 
since determining the $S$-matrix in a general system with multiple
particles is a high non-trivial task, it is valuable to pursue
alternative ways of approaching the virial expansion.

Approximations or assumptions are unavoidable at some point.
The traditional strategies have been either to limit the Hilbert space
allowed for the variational many-body wave functions or to design
schematic Hamiltonians aiming for specific features. The latter 
approach requires great care and physical intuition to retain the 
necessary features of the Hamiltonian that will accurately describe
the phenomena under study. An approach along this second line of 
reasoning is the use of harmonic Hamiltonians where interaction
terms are replaced by harmonic oscillators. This is extremely
convenient from a computational point of view as many aspects
become analytically addressable for both fermionic and 
bosonic systems \cite{brosens97,magda00,brosens,tempere}.

The replacement of one- and two-body terms in the Hamiltonian by
harmonic forms leaves the problem of determining the parameters
of the harmonic Hamiltonian according to given criteria. Here one 
must again take guidance from physical properties and aspects of 
the system that are crucial for the system under study. Recently,
we formulated an approach that explores the $N$-body problem in
an external parabolic confining potential by
fixing the two-body interactions to the properties of an exactly 
solvable problem in the same geometry \cite{arm11}. The two-body 
information needed is the energy eigenvalues and structural 
properties of the wave function such as radial averages.

The example studied in Ref.~\cite{arm11} was short-range interacting
particles in a harmonic trap for which the two-body problem can 
be exactly solved in the zero-range limit \cite{busc98}. Within 
the field of cold atomic gases, this solution was subsequently confirmed
by different experimental groups \cite{stoferle06}. The model studied
in Ref.~\cite{busc98} has since been used as a starting point in 
both nuclear and cold atomic gas physics \cite{haxton02}. Recently,
the harmonic approximation with parameters fixed to exact two-body 
properties have been applied to particles
that interact via dipole forces \cite{arm10,vol12a,dip12,wang06,pikovski11}
and shown to accurately reproduce numerical few-body results for 
moderate-to-strong dipole strengths \cite{shih09,klawunn10,baranov11,vol11}.

This harmonic method has also been extended to the thermodynamical
regime at finite temperature \cite{magda00,brosens97,arm12a}, where
it has been shown using a path-integral formalism that the 
canonical partition function for given particle number can 
be obtained \cite{brosens97,lemmens1999,tempere,klimin2004,brosens2005}. 
Here we consider an alternative approach that uses exact 
diagonalization of the 
Hamiltonian and subsequent calculation of the relevant 
degeneracies in the energy spectrum for a given number
of particles \cite{arm12a}. 
This is in contrast to the usual approximation
using the grand partition function where only the average particle
number is conserved.  
The method applies to both identical bosons and
fermions as well as distinguishable particles and combinations of all
these possibilities \cite{arm11,arm12a,vol12a}.  
The difficult part is to find the degeneracies
of the $N$-body spectrum for the specified symmetries required by
quantum statistics.  The energies themselves are easily found from the
harmonic oscillator solutions.

Here we consider the virial expansion within the 
harmonic Hamiltonian approximation for identical fermions. 
This paper
is a natural extension of Ref.~\cite{arm12a} where the 
thermodynamics of small to moderate size systems of fermions and 
bosons was considered by direct computation of the partition function.
This requires a numerically efficient determination of the level 
degeneracies
which is, however, only possible up to moderate particle numbers
($N\sim 20$) even in a harmonic model. The virial expansion is 
usually a rapidly converging series and we therefore expect to be able
to compute coefficients within the harmonic approach. However, there are
some subtleties with the convergence of the coefficients that must be 
carefully handled. Therefore we focus almost exclusively on the formal 
development of the virial expansion within the harmonic approximation.

A motivation for our work is the recent investigation of universal 
thermodynamics within cold atomic gas experiments \cite{ufermi} where the virial 
expansion has been succesfully applied \cite{virial}. However, the 
expansion is general and is applicable to other fermionic systems.
While the typical condensed-matter and cold atom fermion systems have two
internal (spin or hyperfine spin) components, we consider the case
of single-component fermions here in order to keep the formalism simple
while still retaining the full quantum statistical properties of a 
Fermi system. Multi-component fermionic and bosonic systems will be 
considered in subsequent studies.

The purpose of the present paper is to formulate and explore the harmonic method 
and the virial expansion to prepare for future applications to systems in 
both cold atomic gases, nuclear physics, as well as condensed-matter systems.  
The paper is organized as follows; We describe the ingredients of the method in Sec.~\ref{method} 
The numerical illustrations follow in Sec.~\ref{numerics} for the lowest virial
coefficients. In Sec.~\ref{summary} we summarize and provide an outlook for future directions
of interest.

\section{Theoretical description}\label{method}
We first present general definitions of the crucial ingredients. Then
we apply the formulation to the results of a system of $N$ particles
described by a coupled set of harmonic oscillator potentials.  The
approach to the large temperature limit is finally modified to exhibit the
behavior corresponding to the correct high energy spectrum.

\subsection{Basic definitions}
The classical virial expansion is an expansion of the equation of
state of a gas of identical particles, usually in powers of the
number density $\rho$, see e.g. \cite{thie85,onne01,urse27,hill56}:
\begin{equation}
p=k_BT\rho (1 + B_2(T) \rho + B_3(T) \rho^2 + \dots),
\label{clas}
\end{equation}
where $p$ is the pressure, $k_B$ is Boltzmann's constant, $T$ is the
temperature, and the $B_i$'s are the virial coefficients of the
expansion.  In the classical expansion, they are related to the
intermolecular or interatomic potentials of $i$ interacting particles.
The advantage of the expansion is that it reveals deviations from
ideal gas behavior by examining just the few-body aspects of the
system.  The leading term is then the ordinary ideal gas expression
for a non-interacting system. The second term in the classical
expansion in three dimensions is given by the $B_2$-coefficient
\begin{equation}
B_2 = -\frac{1}{2}\int[\exp(-\beta V_{12}(r))-1]d^3r,
\label{e33}
\end{equation}
where $V_{12}$ is the inter-particle potential depending on the
relative coordinate $r$, and $\beta=1/k_BT$.  The integral in
Eq.~\eqref{e33} is known as a configuration integral, and from it one
can see that this coefficient only converges for potentials which
decay faster than $1/r^3$.  The convergence in two dimensions is
correspondingly only achieved when $V$ decays faster than $1/r^2$.
The general coefficient is
\begin{eqnarray}
B_i&=&-\frac{i-1}{i!V}\\
&&\sum\textrm{(all independent i-cluster integrals)},\nonumber
\end{eqnarray}
where $V$ is the volume of the system. The i-cluster integrals are all the 
independent clusters containing the i-particles, first presented graphically 
by \cite{maye40}.  A cluster in this context can be visualized graphically by 
thinking of $i$ numbered circles with lines connecting them symbolizing the 
interaction between those particles.  These lines can be drawn in many 
different ways, the only requirement for the $i$-cluster is that all $i$ 
atoms or molecules must be connected to at least one other member of the 
system.  The quantum mechanical version of the cluster expansion was 
developed at around the same time by \cite{kahn38}.

For a system of quantum mechanical particles with Fermi or Bose
statistics, the expansion is most commonly performed in the fugacity,
$z=\exp{(\beta\mu)}$, of the system, where $\mu$ is the chemical potential
of the $N$-body system. The grand canonical partition function, $\mathcal{Z}$, is 
written as an expansion in $z$
\begin{equation}
\mathcal{Z}=1+zQ_1+z^2Q_2+\dots,
\end{equation}
which translates into an expansion for the grand
thermodynamic potential, $\Omega$:
\begin{equation}
\Omega=-k_BTQ_1[z+b_2z^2+b_3z^3+\dots].
\label{e41}
\end{equation}
The first virial coefficients can explicitly be written:
\begin{eqnarray}
b_2 &=& (Q_2-Q_1^2)/Q_1  \label{e47} \\
b_3 &=& (Q_3-Q_1Q_2+Q_1^3/3)/Q_1 \label{e52} \\
b_4 &=& (Q_4-Q_1Q_3  + Q_1^2 Q_2  + Q_2^2/2 - Q_1^4/4)/Q_1  \;, \label{e53}
\end{eqnarray}
where $Q_N$ is the canonical partition function for $N$ particles of
the proper symmetry, that is
\begin{equation}
Q_N = \sum_j g_j^{(N)}\exp(-\beta E_j^{(N)}) \;,  \label{e54}
\end{equation}
where $g_j^{(N)}$ and $E_j^{(N)}$ are the degeneracy and energy of the
$j$'th state of the $N$-body system. Thus $Q_N $ can be calculated solely
from the energy spectrum of the $N$-body system, and all thermodynamic
quantities can then be obtained from $\Omega$
in Eq.~\eqref{e41} to the order desired. The classical virial 
expansion, Eq.~\eqref{clas}, can be recovered from the quantum version
introduced above by 
using $\langle N\rangle=-\tfrac{\partial\Omega}{\partial\mu}|_{\textrm{T,V}}$
and $\Omega=-pV$ (see for instance Ref.~\cite{reichl1998} where the relation of 
$B_i$ and $b_i$ is also discussed). In the present case
we have an external trap and the volume must be suitable translated
into parameters of the trap before making detailed comparisons
to experiments or other studies \cite{hu2010}.

In practical calculations, it is more convenient to consider the
difference between interacting and non-interacting systems.  We then
consider the differences $\Delta Q_n=Q_n-Q_n^{(1)}$ and $\Delta
b_n=b_n-b_n^{(1)}$, where the superscript (1) denotes a
non-interacting system having the same $N$-body fugacity $z$.  We can
then re-write Eq.~\eqref{e41} as
\begin{equation}
\Omega=\Omega^{(1)}-k_BTQ_1[\Delta b_2z^2+\Delta b_3z^3+\dots] \; ,
\end{equation}
where $\Omega^{(1)}$ is the grand thermodynamic potential of the
non-interaction system with the same fugacity.  The differences of the
virial coefficients in Eqs.~\eqref{e47}-\eqref{e53} become
\begin{eqnarray}
\Delta b_2 &=& \Delta Q_2/Q_1 \label{e47a} \\
\Delta b_3 &=& \Delta Q_3/Q_1-\Delta Q_2  \label{e52a} \\
\Delta b_4 &=& \Delta Q_4/Q_1 - \Delta Q_3 + Q_ 1 \Delta Q_2 \nonumber\\
&+& (Q_{2}^{2}-(Q_{2}^{(1)})^2)/(2Q_1) \;. \label{e53a}
\end{eqnarray}

These definitions and expressions are general and now applicable to a
specified set of potentials producing the partition functions $Q_n$
and the corresponding (differences of) virial coefficients.

\subsection{Harmonic approximation}
The one- and two-body interactions for the $N$-body system are
approximated by second order polynomials in Cartesian coordinates.
The Hamiltonian is a sum of similar terms from each spatial dimension, 
$H=H_x+H_y+H_z$. In the present work we consider the two- and three-dimensional 
cases. For identical particles of mass $m$ the Hamiltonian
of the $x$-direction is
\begin{eqnarray}
 H_x&=& - \frac{\hbar^2}{2m}\sum_{k=1}^{N} \frac {\partial^2}{\partial x_k^2}+
e \frac{1}{8}m\omega_{in}^2\sum_{i,k=1}^{N}(x_i-x_k)^2\nonumber\\
&&+\frac{1}{2}m\omega_0^2\sum_{k=1}^{N} x_k^2+\frac{N(N-1)}{2}V_{sx}  \; ,\label{e80}
\end{eqnarray}
where $x_i$ is the coordinate of particle $i$, $\omega_{0}$ and
$\omega_{in}$ are frequencies of the one and two-body interactions,
and $V_{sx}$ is a constant adjusting the energy to the desired value.
The factor 1/8 in the second term comes from the use of the reduced
mass equal  to $m/2$ and to avoid double counting in the sum. 

The frequencies and the constant shift can be chosen to reproduce certain
properties of a modeled system as discussed in \cite{arm11}.  
How these parameters are
chosen does not affect the computation of the virial coefficients
which therefore are obtained as functions of the parameters in the
Hamiltonian. The method is general and applicable as soon as a
Hamiltonian of the oscillator form is available.  
It is, however, still useful to illustrate by describing the procedure 
for a specific system.  We focus on a system of identical, spin-polarized 
fermions confined in an external trap \cite{arm11} where the fermions interact
via a short-range potential. Due to the Pauli principle, the particles 
cannot interact in the spherical $s$-wave channel, and the lowest non-trivial 
interaction will be odd and of the $p$-wave kind (higher odd partial
wave channels will be neglected). 
In the zero-range limit, the model of Busch {\it et al.}
\cite{busc98} can still be solved for $p$-wave interactions in both
three- \cite{pwave3D} and two-dimensional \cite{pwave2D} traps.
We adjust the interacting frequency, $\omega_ {in}$, to reproduce some
property related to the spatial extension of the correct two-body wave
function (the average square radius in the two-body ground state in the 
trap). The shift, $V_{sx}$, is then added to make sure that the
exact two-body ground state energy is reproduced by the oscillator
potential. Notice that this (constant) energy shift does not influence
thermodynamics in any essential way. We therefore ignore it for 
most of our discussion except for some comments near the end 
of Sec.~\ref{numerics}. 

The accuracy of the harmonic approximation should depend on 
the degree to which the physical two-body potential allows
a quadratic expansion. Naively, this should be the case for
potentials that have a sizable attractive pocket, which is 
true for many molecular potentials that allow a large number
of bound states \cite{peder2012}. An example is the Morse
potential which has been explored in the harmonic approximation 
in Ref.~\cite{tempere}. As mentioned in the introduction, 
for dipolar particles, the harmonic approximation is 
extremely accurate, even in the regime of small dipole moments
when suitable adjustment of the harmonic frequency is 
performed \cite{arm10,vol12a,dip12}. In fact, even when 
the real potential is shallow, the energy can be reproduced
to a few percent accuracy with a careful choice of gaussian
wave function as shown in Ref.~\cite{dip12}.

For typical cold atomic gas setups, one has harmonically trapped
atoms interacting via short-ranged interactions. In the idealized
limit of zero-range interactions, the two-body problem is 
exactly solvable as demonstrated by Busch {\it et al.} \cite{busc98}.
In Ref.~\cite{arm11}, the exact solution was used to fit the 
oscillator parameters that provide the input for the harmonic
approximation. In the strongly-bound limit where a deep 
two-body bound state occurs, this choice of parameters leads
to the same scaling of the energy with particle number that 
is observed in variational approaches \cite{variational,fu2003}.
Also, when the interactions have a diverging two-body
scattering length (the unitarity limit), the two-body
wave function becomes similar to the non-interacting 
wave function in the trap \cite{busc98} and we therefore
expect the harmonic approximation to be very good. These
features are clearly seen in the one-dimensional 
case as discussed in Ref.~\cite{jeremy2012}. As 
discussed in Ref.~\cite{busc98}, the one- and three-dimensional 
cases are very similar. The scalings away from these
limiting cases are similar but not identical to other approaches.
In general, we expect the harmonic approximation to give good
qualitative results for strong interactions, but do not expect
perfect quantitative agreement with variation or numerics.
On the quantitative side, we note that for one-dimensional 
systems, the three-body energy can be reproduced to within 
10 percent in the strongly-bound limit \cite{jeremy2012}. 
We thus estimate similar accuracy on the third virial 
coefficient.
At this point we leave the question of 
how to adjust the two-body parameters of the Hamiltonian and 
proceed with a general discussion for arbitrary parameters.

The solution to Eq.~\eqref{e80} is found by a coordinate transformation which splits
the Hamiltonian into $N$ independent harmonic
oscillators with new coordinates and frequencies related to the normal
modes of the $N$-body system.  For $N$ identical particles two new
normal mode frequencies are produced, that is the external trap
frequency, $\omega_0$, corresponding to the center of mass motion, and
the $N-1$ times degenerate frequency, $\omega_r$, given by
\begin{equation}
\omega_r^2=N\omega_{in}^2/2+\omega_0^2 \; .  \label{e98}
\end{equation}
The degenerate frequencies correspond to different types of 
intrinsic (relative) motion
which for two particles is simply oscillations in the relative
coordinate, but in the general case correspond to different normal modes
of the system. The $N$-body energy spectrum is 
\begin{eqnarray} \label{e82}
 &E_j = E_{cm} + E_{rel} + V_S, V_S = \frac{D}{2} N(N-1)V_{sx},& \\
 &E_{cm} = \hbar\omega_0\left(n_0(j)+\frac{D}{2}\right),
 n_0(j) = \sum_{i=1}^{D} n_{0,i}(j), \label{e84}& \\
 &E_{rel} = \hbar\omega_r(n_r(j) + \frac{D}{2}(N-1))+ V_S,& \label{e86} 
 \\ &n_r(j) = \sum_{i=1}^{D} \sum_{k=1}^{N-1} n_{k,i}(j)  = 
  \sum_{i=1}^{D} n_{r,i}(j),&
\end{eqnarray}
where $j$ is the index of the $N$-body states. $n_{0,i}$ is the number
of excitation quanta in the center-of-mass degree of freedom in the $i$'th
direction, while $n_{k,i}$ is the number of quanta in the $k$'th 
of the $N-1$ degenerate $\omega_r$ modes in the $i$'th direction.
The energy state $j$ is thus given by specifying the number of 
excitation quanta in each normal mode for all dimensions. Here
we have divided into
center of mass and relative contributions since in the equal mass case
studied here these can be separated completely.

\subsection{The lowest virial coefficients}
The partition function for one particle, $Q_1$, is the trivial problem
of one particle in an external harmonic trap of $D$ dimension.  Since it has
no other particles to interact with, the spectrum arises only from
center of mass motion for one particle.  The partition function is in
this case found from Eqs.~\eqref{e54} and \eqref{e84} to be
\begin{equation} \label{e103}
Q_1=\left(\frac{\exp(-\Theta_0/(2T))}{1-\exp(-\Theta_0/T)}\right)^D
 = \frac{1}{2^D}\sinh^{-D}\big(\frac{\Theta_0}{2T}\big)\;,
\end{equation} 
where $\Theta_0=\hbar\omega_0/k_B$.  

The partition function for two particles, $Q_2$, is found from
Eqs.~\eqref{e54}, \eqref{e82}, \eqref{e84}, and \eqref{e86}, and can be
factorized into center of mass and relative contributions.  The center
of mass piece is completely symmetric in all the coordinates, so it
plays no role in determining the overall symmetry of the system.  It
is just a geometric series in $D$ dimensions, and in fact equal to
$Q_1$, since the frequency is that of the external trap.

The relative motion corresponds to the difference between the two
individual coordinates. For fermions this motion then must provide the
antisymmetry of the wave function corresponding to an odd number of
relative oscillator quanta, $n_r(j)$, in Eq.~\eqref{e86}. The energies
are completely specified by the quanta and the related degeneracy is
easily counted for the relative motion of two particles.  To obtain an
analytical, closed solution, it is convenient to consider the
individual Cartesian quanta.  In 2D, one merely needs to keep either
quanta, $n_{r,x}$ or $n_{r,y}$ odd in $x$ or in $y$-directions.  In
3D, there are two possiblities, in that all three quanta are odd or
two are even and one is odd.  With these restrictions the summation in
Eq.~\eqref{e54} leads for $D=2$ to 
\begin{equation}
Q_2=Q_{1}\left(\frac{\exp(-\Theta_r/T)}{1-\exp(-2\Theta_r/T)}\right)^2
 \exp(-\Theta_S/T)  \, ,   \label{e109}
\end{equation}
where $\Theta_S=V_S/k_B$ and $\Theta_r=\hbar\omega_r/k_B$.  For $D=3$,
we find instead
\begin{eqnarray}  \label{e114}
Q_2 &=& Q_1 \exp(-\Theta_S/T) \\  \nonumber  && \times
 \frac{3\exp(-5\Theta_r/(2T))+\exp(-9\Theta_r/(2T))}
 {[1-\exp(-2\Theta_r/T)]^3}
\end{eqnarray}

The next terms involve the partition function, $Q_3$, for the
three-body system.  A completely closed form solution for the
partition function is not found, and we keep the expression as an
energy sum over three-body states.  The center of mass summation is 
again performed analytically, and Eqs.~\eqref{e82} and \eqref{e86} lead 
to the other factors amounting in total to 
\begin{equation}
Q_3=Q_1\exp(-\Theta_S/T)\sum_{l=0}^{\infty} g_l \exp(-(l+D)\Theta_r/T) \;,
\label{e122}
\end{equation}
where the summation over all states is reduced to run over all
integers.  The difficulty is then only to know the corresponding
degeneracy, $g_l$.  This number of states of a given excitation energy
is found by the method described in \cite{arm12a}.  The expression in
Eq.~\eqref{e122} is formally the same in two and three dimensions but
the degeneracy factors differ substantially.  The infinite sum must in
practice be truncated at some level of excitation. Ideally this is
after convergence is reached. However, this depends strongly on the
value of the temperature and we therefore first must decide how large
$T$-values we need to investigate.

The differences between interacting and non-interacting virial
coefficients are found from Eqs.~\eqref{e47a} and \eqref{e52a}.  The
non-interacting partition functions are structurally the same as the
interacting ones, only with $\Theta_r$ replaced by $\Theta_0$, and
$\Theta_S= 0$.  Using the expressions in Eqs.~\eqref{e109}, \eqref{e114},
and \eqref{e122}, we can therefore easily find $\Delta b_2$ and $\Delta
b_3$.  For $\Delta b_2$ in two dimensions we explicitly get
\begin{eqnarray}
\Delta b_2&=&\exp(-\Theta_S/T)\left(\frac{\exp(-\Theta_r/T)}{1-\exp(-2\Theta_r/T)}\right)^2\nonumber\\
&&-\left(\frac{\exp(-\Theta_0/T)}{1-\exp(-2\Theta_0/T)}\right)^2 \,,
\label{e130}
\end{eqnarray}
and a slightly more complicated expression for three dimensions from
Eq.~\eqref{e114}.  In the same way we can of course re-write $\Delta
b_3$ in terms of the expression for $Q_3$ in Eq.~\eqref{e122}, but due
to the lack of closed form it does not provide any further
information.

Notice that the low-temperature limit of the virial coefficients 
above is determined by the value of the shift, $\Theta_S$. From
Eq.~\eqref{e130} we see that $\Delta b_2$ will vanish for $T\to 0$
whenever we have $\Theta_S+2\Theta_r>0$, similarly for the higher
virial coefficients. In the following we will mostly discuss the 
case $\Theta_S=0$, i.e. no shift at all, since this is the relevant 
case for thermodynamics. For generality, we comment briefly on the 
influence of the shift for all temperatures in Sec.~\ref{numerics}.

\subsection{Large-temperature limits}
At high temperature, the interacting system should approach a
non-interacting system as the kinetic energy dominates the potential.
This does not imply that all $\Delta b_i$ must vanish at large $T$,
since deviations in the low-energy spectrum between interacting and
non-interacting systems easily produce different large-temperature
contributions. This can be seen by dividing the sum over
states in the partition functions in low- and high-energy parts. Even
if we assume that the high-energy interacting and non-interacting spectra become equal 
and would thus not contribute to $\Delta b_i$,
the low-energy parts remains different and this yields a contribution 
to the virial coefficient for all
temperatures.  However, such difference must remain finite since it
arises from a finite energy interval.

As stated before, the virial expansion does not work for potentials
that do not vanish at a fast enough rate, so one would think that the
harmonic potential, which does not vanish at all, would cause
problems.  Indeed, if we use the derived expressions all our $\Delta
b_i$ diverge with increasing temperature.  The origin of this problem
is simply that the energy spectrum is obtained for the $N$-body
solution of a temperature independent Hamiltonian adjusted to
reproduce ground state properties.  This does not account for the
influence of temperature on the effective interactions and, in turn, on
the energy spectrum. This may also be expressed in terms of an
excitation energy dependence of the effective interaction as seen for
example in the variation of the mean-free-path.  Since excitation
energy on average can be related to temperature these formulations are
equivalent.

To pin-point the problem and subsequently cure it we start with
$\Delta b_2$.  In the limit of high temperature, Eq.~\eqref{e130} can
be expanded to leading orders in $T$ to give
\begin{eqnarray} \label{e132}
\Delta b_2(T\to\infty)=  T^2\left(\frac{1}{4\Theta_r^2}-\frac{1} 
{4\Theta_0^2}\right) - T \frac{\Theta_S}{4\Theta_r^2}  \;
\end{eqnarray}
Equivalently we find for three dimensions that
\begin{eqnarray} \label{e133}
\Delta b_2(T\to\infty) &=&  T^3\left(\frac{1}{2\Theta_r^3}-\frac{1}
{2\Theta_0^3}\right) - T^2 \frac{\Theta_S}{2\Theta_r^3} \\ \nonumber  
 &+& \frac{T}{4}\bigg( \frac{\Theta_S^2}{\Theta_r^3}
 +  \frac{15}{\Theta_r} -\frac{15}{\Theta_0}\bigg) \;
\end{eqnarray}
Obviously $\Delta b_2$ diverges as $T$ to the power of the dimension
$D$, unless of course interacting and non-interacting frequencies are
precisely equal.  The next orders are independent of $T$ or vanishing
with increasing $T$.

The unphysical divergence is here clearly seen to originate from the
difference between the energy spacing found by a fit to ground state
properties and the spacing for a non-interacting system.  The proper
thermodynamics requires a correct spectrum for all energies or
temperatures.

In order for $\Delta b_2$ to be finite at large temperatures,
$\omega_r$ must approach $\omega_0$.  From Eq.~\eqref{e98} this seems
most reasonably achieved by a vanishing interaction frequency
$\omega_{in}$.  Furthermore, Eqs.~\eqref{e132} and \eqref{e133} imply
that the energy shift adjusted to fit the ground state energy also
must vanish in the large-temperature limit.  To get finite $\Delta
b_2$ we introduce cutoff functions $F(T)$ and $G(T)$ into
$\omega_{in}$ and $V_S$:
\begin{equation}
\omega_r^2=\frac{N}{2}F(T)\omega_{in}^2+\omega_0^2 \;,
 \;\; V_S \rightarrow V_S G(T) \; .
\label{e147}
\end{equation} 
Without further constraints, there is a great deal of freedom in the
form of $F(T)$ and $G(T)$, but they must satisfy the conditions of
approaching zero respectively as $1/T^D$ and $1/T^{D-1}$ at high
temperatures.  The simplest such functions 
\begin{equation}
F(T)=\left(\frac{T_0}{T+T_0}\right)^D \;,\;\,
G(T)=\left(\frac{T_0}{T+T_0}\right)^{(D-1)} \;,
\label{e152}
\end{equation}
where $T_0$ is a cut-off parameter which indicates at what temperature
the spectrum continuously should shift to the non-interacting
spectrum.  These functions regularize the high temperature behavior
and now $\Delta b_2$ approaches a constant at high temperatures.  The
limits in the different dimensions can be found from Eqs.~\eqref{e132}
and \eqref{e133} by use of Eqs.~\eqref{e147} and \eqref{e152}, that is
\begin{eqnarray} \label{e132a}
\Delta b_2(T\to\infty)=  - \frac{N}{8}
 \frac{\omega_{in}^2}{\omega_{0}^2}\frac{(k_B T_0)^2}{(\hbar\omega_{0})^2}
  - \frac{V_S k_B T_0}{4(\hbar\omega_{0})^2} \;,
\end{eqnarray}
for two dimensions and for three dimensions we get
\begin{eqnarray} \label{e133a}
\Delta b_2(T\to\infty) = - \frac{3N}{8} \frac{\omega_{in}^2}{\omega_{0}^2}
 \frac{(k_B T_0)^3}{(\hbar\omega_{0})^3}
  - \frac{V_S (k_B T_0)^2}{2(\hbar\omega_{0})^3} \;,
\end{eqnarray}
where $N=2$ for this second virial coefficient.  The high temperature
limit is then determined by the interaction frequency, ground state
energy shift, and the cut-off temperature.  The two terms have opposite
sign since $V_S$ is negative, and initially introduced to balance the
zero-point energy of the oscillator such that large $\omega_{in}$
is correlated with a large negative $V_S$. 

Once the $\Delta b_2$ is regularized we turn to $\Delta b_3$ which is
more complex as seen in Eq.~\eqref{e52a}. It consists of two terms, the
difference between interacting and non-interacting partition functions
for three particles and $\Delta Q_2=Q_1\Delta b_2$.  This latter term
contains the already regularized factor, $\Delta b_2$, but $Q_1$ in
Eq.~\eqref{e103} also diverges as $\rightarrow T^D/\Theta_0^D$ at high
temperature.  Therefore $\Delta b_3$ can only remain finite in the
limit when the difference $(Q_3-Q_3^{(1)})/Q_1$ diverges precisely as
$Q_1$.  

However, this is not even sufficient because the $T^D$ divergence from
$Q_1$ leads to divergence of the terms in $\Delta b_2$ vanishing as
$1/T$ in two dimensions and as $1/T$ and $1/T^2$ in three
dimensions. These terms all must be cancelled by corresponding
diverging terms in $(Q_3-Q_3^{(1)})/Q_1$.  These conditions seems
impossible to meet, but nevertheless this miracle seems to occur.  We find
that $\Delta b_3$ also is regularized with precisely the same cut-off
function as used for $\Delta b_2$ and this occurs in both two and
three dimensions.  Problems with divergences in $\Delta b_3$ have been 
reported by other authors \cite{liu10a}.  They resolved it by 
separating
the system into a 2+1 subsystem in which the divergence was then 
removed.  While we do not need to do this to obtain a finite $\Delta 
b_3$, the underlying deep reason for this is still unclear to us.

The form of the cut-off in Eq.~\eqref{e147} can be changed in many
ways. The transition can set in at different temperatures and be more
or less fast.  We have chosen to use only one parameter, $T_0$, but we
tested with another functional form:
\begin{equation}
F(T)=(1-\exp(-T_0/T))^D \;,
\label{e164}
\end{equation}
which has the same high temperature behavior.  Both $\Delta b_2$ and
$\Delta b_3$ are again regularized in the large-temperature limit.
The values of $\Delta b_2$ and $\Delta b_3$ are both larger in
magnitude at all temperatures simply because this cut-off function is
larger for all $T$.

Instead of the temperature cutoff in Eq.~\eqref{e147}, we could choose a
cutoff in excitation energy.  This might at first appeal as being more
physically reasonable since this directly amounts to changing the
$N$-body spectrum at high excitation energy to the non-interacting 
$N$-body spectrum. 
However, this conclusion is
rather shaky.  Completely different configurations specified by sets
of quantum numbers can give precisely the same energy.  This is easily
seen in a single particle picture by comparing (i) states of the same
total energy arising from a few particles at very high-lying levels
and all others in the lowest possible levels with (ii) the opposite
where all particles are in levels of intermediate energy.

Thus, a given excitation energy already corresponds to an average over
many configurations very similar to the configurations occupied for a
given temperature.  In any case we investigated a cutoff in excitation
energy instead of temperature, that is expressions of similar
simplicity as those in Eq.~\eqref{e152}:
\begin{equation}
F(E)=\left(\frac{E_0}{E+E_0}\right)^D, \;
G(E)=\left(\frac{E_0}{E+E_0}\right)^{(D-1)},
\label{e152b}
\end{equation} 
where $E$ is the excitation energy.  This function is inserted in
Eq.~\eqref{e147} in the place of $F(T)$, and when the partition
function is being calculated as the sum over states, the step in
energy taken between states is now dependent on the position in the
spectrum.  The mean field spacing is reached for energies higher than
$E_0$.  Implementing this cutoff successfully regularizes $\Delta
b_2$, but to remove the divergence for $\Delta b_3$, it is necessary
to use a constant, $E_0$, different from that of $\Delta b_2$.  No
obvious relation is found and generalizations to higher $\Delta b_i$
seems to be at least very impractical.

\section{Numerical results}\label{numerics}
The virial coefficients can now be calculated numerically. In the
model they are completely determined from trap frequency $\omega_0$,
interaction frequency $\omega_{in}$, shift energy $V_S$, and cut-off
function and related parameters $T_0$.  We use the trap frequency as
energy unit which implies that results for any other value of
$\omega_0$ can be obtained by scaling all energies
$\hbar\omega_{in}$, $k_BT_0$, and $V_S$ by $\hbar\omega_0$.  The
energy shift, $V_S$, is introduced to adjust to the correct energy and
has no effect on eigenvalues and corresponding wave functions.  It is
strongly dependent on which model is approximated.  We shall therefore
first investigate the general dependencies on $\omega_{in}$ and $T_0$
which in turn can be related to specific models.  Afterwards we shall
separately investigate the dependence on the shift.

\begin{figure}[ht!]
\includegraphics[width=0.5\textwidth]{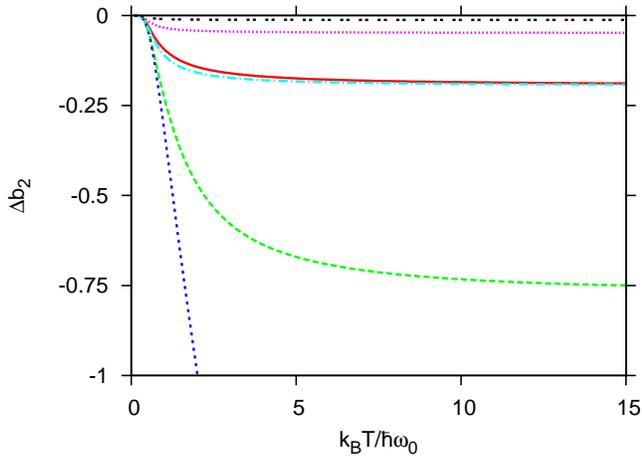}
\caption{$\Delta b_2$ in 2D as a function of temperature with
  a model space of 160 oscillator shells and zero energy shift.  
  The cutoff used was the function in Eq.~\eqref{e152}.
  The values of $T_0$ and $\omega_{in}$ (in units of $\omega_0$)
  from bottom to top are $(T_0,\omega_{in})=(0.51,5.0)$, $(0.25,5.0)$,
  $(0.13,5.0)$, $(0.25,2.5)$, $(0.13,5.0)$, and $(0.062,2.5)$.
  The $T_0$ values are $\omega_{in}/2\pi^2$ and twice and half that value.}
\label{f184}
\end{figure}

\subsection{Virial coefficients}

The cut-off parameter is essential for the behavior of the expansion
coefficients.  The size of an appropriate value can be estimated by
inspection of the effect it is designed to simulate.  The first
excited state appears at an excitation energy of $\hbar \omega_{r}$
which is a single quasi particle excitation. Therefore $\hbar
\omega_{r}$ represents a shell gap to be overcome by thermal
excitations.  This gap is known to wash out at a critical temperature,
$T$ given by $k_B T 2\pi^2 \approx \hbar \omega_{r}$, see
\cite{jen73,boh75}.  This surprisingly large factor, $2\pi^2 \approx
19.7$, suggests a rather small relative value of $T_0$ proportional to
$\omega_{r}$, that is $k_B T_0 \approx \hbar \omega_{r}/(2\pi^2)$.

\begin{figure}
\includegraphics[width=0.5\textwidth]{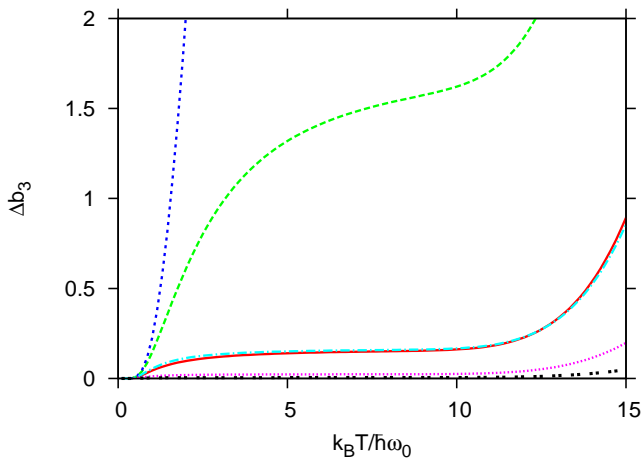}
\caption{Same as in Fig.~\ref{f184} for $\Delta b_3$ in 2D. The 
parameters are also the same but must be read from top to bottom.}
\label{f190}
\end{figure}

To illustrate the dependence we show $\Delta b_2$ in Fig.~\ref{f184}
as function of $T$ for different interactions and cut-off values.  We
see the general behavior of a second order increase from zero at
$T=0$, and the smooth curvature before bending over to reach the
saturation value.  The expansion coefficient is a rather strongly
increasing function of both interaction frequency and cut-off
parameter.  The functional form of the cut-off function in
Eq.~\eqref{e152} implies that the saturation value as well as
saturation temperature both depend rather strongly on $T_0$.

The overall behavior must be understood in the model even at
uninterestingly large temperatures. We continue to show $\Delta b_3$
in Fig.~\ref{f190} for for two dimensions for the same set of
parameters as in Fig.~\ref{f184}.  Qualitatively the same behavior
except for the overall opposite sign.  However, the temperature
dependence is faster, the saturation values are larger, as seen for
the small interaction frequency with the small $T_0$.  For higher
values we observe a tendency to form a flat region which quickly
becomes an increasing function.

At low temperature, the coefficients vanish since both the interacting and 
non-interacting partition functions go to unity at low temperature (in the
abscence of any energy shift).  This might seem odd since our $\Delta b_i$'s are
the difference between a non-interacting and interacting system, which should
increase at low temperatures when the interactions are more significant
compared to the kinetic energy.  Since both of the partition functions are small
at low temperature, the only signature we see of this is that the relative
differences of the partition functions, for example $(Q_2-Q_2^{(i)})/Q_2^{(i)}$, does increase.

\begin{figure}[h!]
\includegraphics[width=0.5\textwidth]{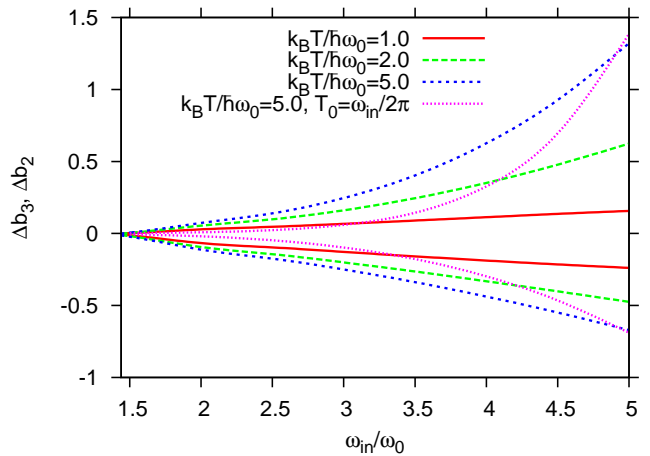}
\caption{ Virial coefficients $\Delta b_2<0$ and $\Delta b_3>0$ as a function
of interaction frequency $\omega_{in}$ for three different temperatures. 
Notice 
that from Eq.~\eqref{e98} we always have $\omega_{in}>\sqrt{2}\omega_0$.
On this plot, $k_BT_0/\hbar\omega_0=0.25$ for all curves except the short-dotted 
line which has  $k_BT/\hbar\omega=5.0$ and $k_B T_0=\hbar\omega_{in}/2\pi^2$ (adjusted 
at end point on the horizontal axis).}
\label{f200}
\end{figure}

\begin{figure}[h!]
\includegraphics[width=0.5\textwidth]{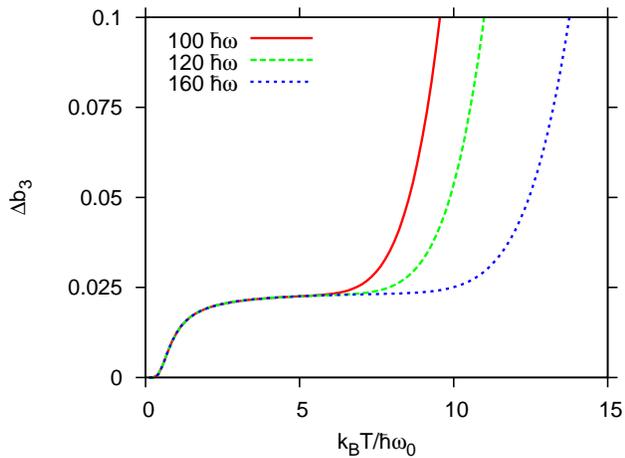}
\caption{ $\Delta b_3$ in 2D as a function of temperature for three
  different model space sizes.  The cut-off function is that given in Eq.~\eqref{e152}.  
  The other paramters are $k_BT_0 = 0.25\hbar\omega_0$ and
  $\omega_{in} = 2.5\omega_0$}
\label{f198}
\end{figure}

The effect of the interaction frequency is plainly to increase the
virial coefficient, which could be seen in Figs.~\ref{f184} and
\ref{f190}.  A more precise dependence on interaction frequency can be
seen in Fig.~\ref{f200}.  The coefficients vanish for small
$\omega_{in}$, as then there is no difference between the non-interacting
and interacting system, but then increase rapidly, especially after
$\omega_{in}>\omega_0$.  The increase is more dramatic at higher
temperatures, which are closer to the saturation value.  It does
appear, however, that for any temperature the behavior of the
coefficients is faster than linear.
\subsection{Hilbert space and cut-off function}

The apparent lack of saturation for $\Delta b_3$ at high temperatures
is a very unsatisfactory feature. Fortunately, it seems to be an effect 
of the
model space truncation at high excitation energies in the calculation
of the partition function.  This happens because of the subtle nature
of the cancellation which removes the divergence and leads to
saturation.  The piece, $Q_1$, in $\Delta Q_2$ in Eq.~\eqref{e52a}
contains all states to infinitely high excitation energies since it is
calculated analytically.  This piece eventually overwhelms the $\Delta
Q_3$ term in Eq.~\eqref{e52a} which is calculated numerically and
consequently arises from a truncated energy spectrum.  

We demonstrate this in Fig.~\ref{f198} where we show $\Delta b_3$ for
three different model space sizes.  The higher energies we include in
the numerical calculation, the larger is the region of the flat
saturation interval, and the larger temperatures before the divergence
sets in.  This is very reassuring allowing us to ignore the unphysical
region of all temperatures above the flat region.

\begin{figure}
\includegraphics[width=0.5\textwidth]{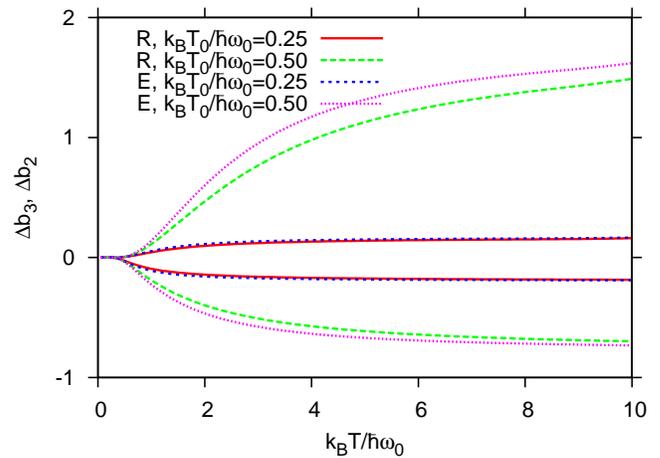}
\caption{ Virial coeffecients in 2D as a function of temperature for
  different cut-off functions. The rational function cutoffs, Eq.~\eqref{e152}, 
  are labeled with an 'R', and the exponential cutoff,
  Eq.~\eqref{e164}, with an 'E'.  $\omega_{in}=2.5\omega_0$ was used for all of
  the curves.}
\label{f204}
\end{figure}

One uncertainty in the method to recover the high-energy
non-interaction limit is the function describing the disappearance of
shell effects.  In Fig.~\ref{f204} we illustrate this dependence of
the virial coefficients by results from use of different cutoff
functions, that is the rational expression, Eq.~\eqref{e152}, and the
exponential functon, Eq.~\eqref{e164}.  The virial coefficients using
the exponential cutoff are larger for all temperatures before finally
merging into the same high temperature limit.  This is due to the
larger cut-off function at all temperatures which leaves the reduction
to take place somehwat faster at larger temperatures although the same
final saturation is reached for $T$ much larger than $T_0$.

\begin{figure}[h!]
\includegraphics[width=0.5\textwidth]{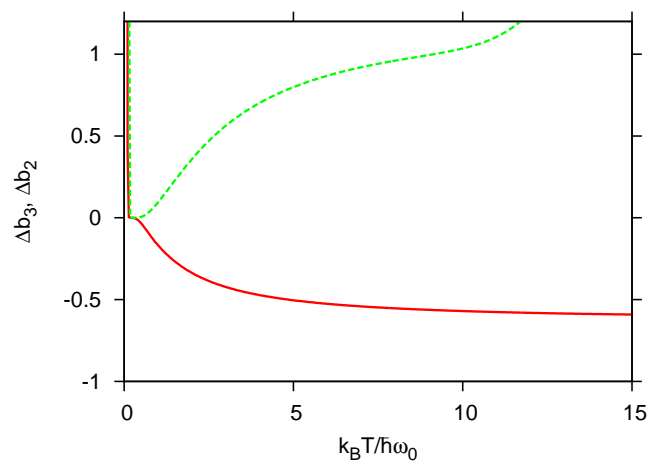}
\caption{ $\Delta b_2$ (lower solid line) and $\Delta b_3$ (upper dashed line)
  in 2D as a
  function of temperature examining the effect of the energy shift.
  The coeffecients behave similar to what was shown
  previously except at very small temperature.  For this plot,
  $\omega_{in}=2.0\omega_0$ with $V_s=-5.81\hbar\omega_0$, and $k_BT_0=0.5\hbar\omega$.}
\label{f206}
\end{figure}

\subsection{Effect of the energy shift}

The energy shift in the Hamiltonian has a different effect on the
virial coefficient.  We illustrate by the examples in Fig.~\ref{f206}.
The large temperature limit remains finite by construction as shown
explicitly in energies Eq.~\eqref{e132a} for one case.  Otherwise the
behavior at large $T$ is very similar to that of zero shift, where
convergence is essential or at least a flat region at high $T$ is
necessary.  The energies enter in the partition function in the
exponent.  Contributions disappear from energies much higher than the
temperature.

However, at very small temperatures the same contribution from the
shift can produce unphysical results.  This is seen at very low $T$ in
Fig.~\ref{f206} where a narrow and large peak is present in both
virial coefficients. The numerical reason is obvious since the
negative shift divided by a small temperature produce a very large
value.  This occurs for temperatures much smaller than $\omega_0$ and
much before the  statistical treatment is meaningful.

The negative shift is not necessarily sufficient for a divergence at
low temperatures.  The shift energy must completely eliminate the
zero-point energy, and thus for polarized fermions $V_S<-D N
\hbar\omega_r$.  The parameters in Fig.~\ref{f206} provide a shift
large enough in magnitude to give a spike towards large positive
values at very low temperatures.  This changes the sign of $\Delta
b_2$, as the term containing the shift eventually (small $T$) becomes
larger than the non-interacting term in Eq.~\eqref{e130}.

For $\Delta b_3$, the singularity at zero temperature comes at a
slightly higher temperature since the shift energy is three times
larger than for two particles.  Again this occurs for temperatures
smaller than those allowing a statistical treatment. 

\begin{figure}[h!]
\includegraphics[width=0.5\textwidth]{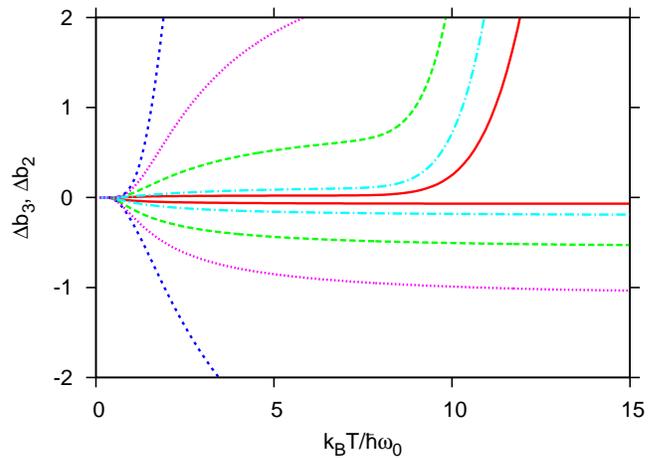}
\caption{The virial coeffcicients in three dimensions, 
  $\Delta b_2<0$ (lower lines) and $\Delta b_3>0$ (upper lines), as function
  of temperature. The cut-off temperatures, $T_0$, and $\omega_{in}$ is varied among the 
  different curves. On the $\Delta b_2$ side (below zero) from bottom to
  top it is $(k_BT_0/\hbar\omega_0,\omega_{in}/\omega_0=(1.00,2.5)$, 
  $(0.50,3.5)$, $(0.50,2.5)$, $(0.50,1.5)$, and $(0.25,2.5)$. For 
  $\Delta b_3$ (above zero), the lines have the parameters just listed 
  but in the inverse ordering.
  For this
  plot, there is no energy shift $V_S=0$ and a model space of 160
  shells is used.}
\label{f215}
\end{figure}

\subsection{Three dimensions}
We have so far only shown results for two dimensions. The method is,
however, as applicable in three dimensions where the coefficients look
qualitatively similar to those in two dimensions, see Fig.~\ref{f215}.
We notice the increase from zero at $T=0$, then a decreasing derivative
resulting in saturation or in a tendency towards saturation, and finally the
divergence at large temperature for $\Delta b_3$ due to mismatch
between the numerical $Q_3$ and the analytical $Q_1$.  The results are
more sensitive to the cutoff parameter because higher powers enter in
two than in three dimensions, as seen for example by comparing
Eqs.~\eqref{e132a} and \eqref{e133a}.  Indeed, if one uses $T_0=\omega_{in}/(2\pi^2)$, then the magnitude of the coefficients is quite small, of order $10^{-2}$.  Also, the truncation effect in
$\Delta b_3$ is more sensitive to the cutoff parameter and can begin
at comparatively low temperatures when compared to 2D, see
Fig.~\ref{f190}.

The virial expansion is most efficient when the size of the
coefficients decrease with the order. Therefore, a requirement of
$\Delta b_3\leq \Delta b_2$, leads to a condition on the maximum size
of $T_0$.  This demand is more restrictive for three than for two
dimensions.

\section{Summary and outlook}\label{summary}
We have discussed
the virial expansion technique and a quantum mechanical
formulation is sketched from an analogous classical expansion. We
apply the formulated method to a harmonic
approximation to the $N$-body problem for identical fermions. 
A key step in this approach is the adjustment of the harmonic 
one- and two-body parameters to pertinent properties of the 
corresponding two-body problem that holds information about the 
exact interaction which is approximated by a harmonic form. Once these
are obtained, the resulting $N$-body Schr{\"o}dinger equation may
be solved exactly and the spectrum can be used to compute the 
partition function. The
second and third order virial expansion coefficients are obtained by
direct calculation of the two and three-body partition functions. 
Here we are interested in the details of the formal development 
of a virial expansion and we therefore vary the parameters of the 
harmonic interaction terms freely to study the behavior. The mapping
to realistic two-body properties is straightforward.

The virial expansion can be reformulated in terms of deviations between
non-interacting and interacting systems.  Importantly, the virial coefficients
have an unphysical divergence for large temperatures. It arises in the
formulation because the increasing temperature populates higher and
higher excited states. Their average properties can be very far from
the ground state properties, and eventually the results should
resemble those of the non-interacting system where the kinetic energy
is decisive.  This is achieved by modifying the energy spectrum by
a function of temperature smoothly connecting the low-temperature ground state 
dominated and high-temperature non-interacting spectra. 

To achieve the goal of removing the divergence, the modification
function must reduce the initial interaction frequency to zero by a
large-temperature behavior where the power of the temperature is equal
to the spatial dimension of the system.  The divergence is removed
from the second order expansion coefficient and it turns out that
the same modification function removes the divergence from
the third order term. This result is highly non-trivial since the
third order divergence is of a very different origin from that of
second order. This is emphasized by an attempt to use an energy (in
contrast to temperature) dependent modification function which removes
the second order divergence but requires an additional adjustment to
remove the third order divergence.  The temperature modification is
then the more promising approach for applications where higher orders have to
be calculated.

We find that the critical temperature value describing where the 
adjustment of the spectrum should take place and the rate of
the modification should be about $2\pi^2$ smaller than the two-body
interaction frequency.  This is an analogy to the smearing of
shell effects by temperature in an $N$-body finite system described by
its single-particle spectrum. This function is exponential and the
rate is precisely the single-particle frequency divided by
$2\pi^2$. In our case the modification function is chosen to have a
rate of change of the $N$-body energy spectrum which is a rational 
function of temperature to a power that depends on dimension. 
However, the modification takes
place on the energies appearing in the exponent of the partition
function.  The precise shape of the temperature modification function
is not essential for the overall properties unless one consideres extreme
cases. A sensible choice of modification function that regularizes the 
divergence will thus yield a formalism that can be adjusted to 
low-energy properties and make further predictions for the many-body 
problem.

While we have focused on two- and three-dimensional systems in the 
current presentation, a good testing ground for the formalism discussed 
would be some of the exactly solvable models that are known in 
one-dimensional systems \cite{sutherland04}. A good example 
is the $N$-boson problem with zero-range interactions studied 
by Lieb, Linniger, and MacGuire \cite{lieb63a} for which the 
harmonic approximation can be applied \cite{jeremy2012}, or to 
dipolar molecules in one-dimensional setups where $N$-body 
clusterized bound states can easily form \cite{dipole1D}. In the 
strong-coupling domain these should be well-described within a
harmonic approximation approach.

In summary, we have demonstrated that the harmonic approximation employed
at the Hamiltonian level gives a divergent set of virial expansion 
coefficients that must be regularized at high temperature. This can 
be done by a careful choice of temperature-dependent spectral 
modification that will render the virial coefficients finite at all
temperatures. The ease of solving the harmonic $N$-body problem 
and subsequently calculating the virial coefficients makes this an
attractive approach to compute many-body properties.

\end{document}